\documentclass{elsarticle}
\usepackage{lipsum}

\usepackage{amssymb, epsf, float, graphicx, graphics, verbatim, amssymb,txfonts,multirow,lscape,setspace}
\usepackage{graphics,array, helvet,bm}
\usepackage{subfigure}

\usepackage{adjustbox}
\usepackage{multirow}
\usepackage{color}

\usepackage[margin=2cm]{geometry}

\usepackage{graphicx}
\usepackage{arydshln}
\usepackage{multirow} 
\usepackage{booktabs}
\usepackage{bm}
\usepackage{epsfig}
\usepackage{hyperref}

\usepackage{animate}

\newcommand{\widthscalefive}{0.145}

\usepackage{etoolbox}
\AtBeginEnvironment{thebibliography}{\linespread{1}\selectfont}

\makeatletter
\def\ps@pprintTitle{%
 \let\@oddhead\@empty
 \let\@evenhead\@empty
 \def\@oddfoot{}%
 \let\@evenfoot\@oddfoot}
\makeatother

\usepackage[ruled,linesnumbered]{algorithm2e}
\SetAlFnt{\scriptsize}

\journal{Neurocomputing}

\title{\hspace{1cm} Deep Networks for Image and Video Super-Resolution}

\author{\hspace{1cm} Kuldeep~Purohit}     
\ead{kuldeeppurohit3@gmail.com}

\author{\hspace{1cm} Srimanta~Mandal}      
\ead{in.srimanta.mandal@ieee.org}

\author{\hspace{1cm} A.~N.~Rajagopalan}    
\ead{raju@ee.iitm.ac.in}

\address{IPCV Lab, Department of Electrical Engineering,
Indian Institute of Technology Madras, TN, India}

\begin{document}
\maketitle

\section{Abstract}

Efficiency of gradient propagation in intermediate layers of convolutional neural networks is of key importance for super-resolution task. To this end, we propose a deep architecture for single image super-resolution (SISR), which is built using efficient convolutional units we refer to as mixed-dense connection blocks (MDCB). The design of MDCB combines the strengths of both residual and dense connection strategies, while overcoming their limitations. To enable super-resolution for multiple factors, we propose a scale-recurrent framework which reutilizes the filters learnt for lower scale factors recursively for higher factors. This leads to improved performance and promotes parametric efficiency for higher factors. We train two versions of our network to enhance complementary image qualities using different loss configurations. We further employ our network for video super-resolution task, where our network learns to aggregate information from multiple frames and maintain spatio-temporal consistency. The proposed networks lead to qualitative and quantitative improvements over state-of-the-art techniques on image and video super-resolution benchmarks.



\section{Introduction}
\label{sec:intro}
Single image super-resolution (SISR) aims to estimate a high-resolution (HR) image from a low-resolution (LR) input image, and is an ill-posed problem. Due to its diverse applicability starting from surveillance to
medical diagnosis, and from remote sensing to HDTV, the SISR problem has
gathered substantial attention from computer vision and image processing
community. The ill-posed nature of the problem is generally addressed by 
learning a LR-HR mapping function in a constrained environment using example HR-LR patch pairs. 

One way is to learn a mapping function that linearly correlates the HR-LR
patch pairs. Such linear functions can be easily learned with few example images as has been practiced by some SR approaches~\cite{yangj,ASDS,Mandal_SRSI}.
But, linear mapping between such patch pairs may not be representative enough to learn different complex structures present in the image. The mapping function would benefit from learning non-linear relationships between HR-LR patch pairs. 
Recent convolutional neural network (CNN) based models are quite efficient for such a purpose, and can be useful in extracting relevant features by 
making deeper models. However, deeper models often face vanishing/exploding gradient issues, which can be partially mitigated by using residual mapping~\cite{he2016deep,ledig2017photo}. Deep residual models has been employed for higher level vision tasks, where batch normalization is generally used for a useful class-specific normalized representation. However, such representation is not 
much useful in low-level vision task such as SR~\cite{lim2017enhanced}.
Most deep CNN based SR models do not make full use of the hierarchical features from the original LR images. Thus, the scope of improvement in
performance is there in effective employment of the hierarchical features
from all the convolutional layers, as has been employed by a residual dense network~\cite{zhang2018residual} using a sequence of residual dense blocks (RDBs). However,
most of these deep networks require huge number of parameters, which becomes a bottleneck in the situation where limited computational resources are available.

Further, most of the networks need to be trained separately for different
scale factors. This issue can be addressed by jointly training the model
for different scale factors, as has been performed by VDSR~\cite{kim2016accurate}.  Moreover, VDSR requires bicubic interpolated version of the LR image as an
input. Processing interpolated images precludes the model from learning direct LR-HR feature mapping. Additionally, passing the high-resolution image and feature-maps through a large number of layers increases the memory consumption as well as computational requirements. 
Another way to consider multiple scale factors is to train a network jointly for different factors by including scale-specific features~\cite{lim2017enhanced}. However, these techniques produce sub-optimal results for higher scale factors like $8$.

Motivated by the performances of residual connections and dense connections, we propose a deep architecture for single image super-resolution (SISR) that consists of efficient convolutional units, which 
we entitle as mixed-dense connection blocks (MDCB). MDCB combines the
advantages of residual and dense connections while subduing their
limitations. The combination improves the flow of information through the network, alleviating the gradient vanishing
problem. In addition, it allows the reuse of feature maps from preceding layers, avoiding the re-learning of redundant features. In order to 
master our network for super-resolving different scale factors, we make use
of weight transfer strategy via scale-recurrent framework. Intuitively, filters learnt for smaller scale factors can be transfered to higher-ones. Sharing of parameters across scales is crucial for efficiently super-resolving by higher up-sampling factors. Our scale recurrence framework built using MDCBs is parametrically more efficient than most of the existing works, enabling our strategy to work with limited resources.

It has been recently found that a result with good RMSE value often fails to
satisfy perceptually, and the converse also holds true ~\cite{blau2017perception}. Thus, to obtain
photo-realistic results, we employ a GAN framework with deep feature loss (VGG)
function in the network, which leads to a second network. These two 
networks enable us to traverse along the perception-distortion curve. The
first network is trained to produce an output with better RMSE score, whereas the second one tries to produce result with better perceptual score. Different weighting schemes of GAN and VGG and pixel reconstruction losses enables us to traverse along the perception distortion curve \cite{blau2017perception} and can be used to reach a desirable trade-off between the two.

We have further employed our network for the task of video SR, where our model super-resolves each frame by aggregating HR information from a local temporal window of LR frames. We demonstrate that the proposed SR framework can approximate the inverted image formation process, while maintaining spatio-temporal consistency and the estimated HR frames are good candidates for representing the ground truth frames of a video. The proposed networks help in achieving better qualitative and quantitative performance against the state-of-the-art techniques on image and video super-resolution benchmarks.

\section{Related Works and Contributions}
\label{sec:relate}
We address the problems of SR for single image and extend it to video. To discuss the related works for each of the problems, we discuss this
section mainly in two parts, corresponding to the single image and video SR.
In each category, there are numerous approaches starting from conventional to deep learning based. Though, our technique is based on deep 
learning, we also brief some of the conventional techniques for completeness.

\subsection{Works on Single Image SR}
Super resolving single image generally requires some example HR images to import 
relevant information for generating the HR image. Two stream of approaches make
use of the HR example images in their frameworks: i) Conventional approaches, and
ii) deep learning based approaches.

\subsubsection{Conventional Single Image SR}
Conventionally, single image SR approaches work by finding out patches similar to the target patch in the database of patches, extracted from example images. However, the possibility of many similarities along with the imaging blur, and noise often makes the problem ill-posed. Different prior information help in 
address the ill-posed nature of the problem \cite{rajagopalan2003motion,rajagopalan2005background,suresh2007robust,bhavsar2010resolution,bhavsar2012range,punnappurath2015rolling,punnappurath2017multi,vasu2018analyzing,purohit2020mixed}.

Natural images are generally piecewise
smooth, hence it is often incorporated as prior knowledge by means of
Tikhonov, total variation, Markov random field, etc.~\cite{Tikhonov_sr,TVSR,MRF2}.
Image patches tend to repeat in the image non-locally, and can provide useful
information in terms of non-local prior to super-resolve an image~\cite{Mairal,ASDS,Glasner,Mandal_SRSI}. The redundancy present in an image 
indicates sparsity in some domain, and can be employed in SR framework
by employing sparsity inducing norm. Here, the assumption is that a target patch
can be represented by combining few patches from the database linearly\cite{yangj,zeyde2012single,ASDS,Mandal_SRSI}.  Different priors can be combined to provide
various information to the framework. For example, sparsity inducing norm 
can be combined with non-local similarity to improve the SR performance~\cite{NCSR,Mandal_SRSI}.

Although, the sparsity based prior works quite efficiently, the linear mapping
of information fails to represent complex structures of image. Here, deep-learning based approaches behave better through non-linear mapping of HR information~\cite{dong2016image,dong2016accelerating,wang2015deep,kim2016accurate,
tai2017image,lai2017deep,MSLapSRN,sajjadi2017enhancenet,tai2017memnet,
lim2017enhanced,zhang2018learning,huang2015single}.

\subsubsection{Deep Learning Based Single Image SR}
Deep learning stepped into the field of SR via SRCNN~\cite{dong2014learning}, which extends the concept of sparse representation using
CNN. CNNs typically contains large number of filters along with series of non-linearities and hence are a better candidate for representing complex input-output mapping than the conventional approaches, and are shown to yield superior results. However, increasing the depth of such architecture increase difficulty in training. Introducing residual connections into the framework along with skip connections and/or recursive convolutions are known to make it less cumbersome. Following such methodologies, VDSR~\cite{kim2016accurate} and DRCN~\cite{kim2016deeply} have demonstrated performance improvement. 
The power of recursive blocks involving residual units to create a deeper network is explored in~\cite{tai2017image}. Recursive unit in conjunction with gate unit acts as a memory unit that adaptively combines the previous states with the
current state to produce a super resolved image~\cite{tai2017memnet}. However, these
approaches interpolate the LR image to the HR grid before feeding it to the network. This technique increases the computational requirement since all the convolution operations are then performed on high-resolution feature-maps. To alleviate this computational burden, networks have been tailored to extract features from the LR image through a series of layers. Towards the end of such networks, up-sampling process is performed to match with the HR dimension~\cite{dong2016accelerating,ledig2017photo}.
This process can be made faster by reducing the dimension of the features going
to the layers that maps from LR to HR, and is known as FSRCNN~\cite{dong2016accelerating}. 

Recent studies show that traditional metrics used to measure image restoration quality do not correlate with perceptual quality of the result. The work of in SRResNet~\cite{ledig2017photo} utilized ResNet~\cite{he2016deep} based network with adversarial training framework (GAN)~\cite{goodfellow2014generative} with perceptual loss~\cite{johnson2016perceptual} to produce photo-realistic HR results. The perceptual loss is further used with texture synthesis mechanism  in GAN based framework to improve SR performance~\cite{sajjadi2017enhancenet}. Though these approaches are able to add textures in the image, sometimes the results contain artifacts. The model architecture of SRResNet~\cite{ledig2017photo} is further simplified and optimized to achieve further improvements in EDSR~\cite{lim2017enhanced}. This is further modified in MDSR~\cite{lim2017enhanced},
which performs joint training for different scale factors by introducing scale-specific feature extraction and pixel-shuffle layers, while keeping rest of the layers common.

\subsection{Works on Video SR}
We briefly discuss some of the related works on video SR including
conventional approaches and recent deep learning based approaches.
\subsubsection{Conventional Approaches}
The seminal work of~\cite{Tsai1984} has enabled various development in multi-frame SR.
Motion between the consecutive frames can be employed in reconstruction based methods~\cite{Schultz_TIP96,Patti_TIP97,Sawhney_ECCV02}. The quality of the super-resolved video depends on
the accuracy of motion estimation of pixels.
In order to maintain the continuity of the result, a back-projection method
was introduced to minimize reconstruction error iteratively~\cite{Irani_JVCI93}. Example-based image SR approaches cannot be directly extended for video by super-resolving each frame separately. Independent 
processing of frames can give rise to flickering artifacts among the frames.
This issue can be mitigated by introducing smoothness prior among the frames~\cite{Bishop_AIS03}. Different spatial-temporal resolutions
of frames can also be combined to super-resolve a video~\cite{Kong_BMVC06}, where a dictionary has been constructed by using scene specific
HR images captured by a still camera. Single image SR has also been exercised
for video SR in frame by frame basis by incorporating inverted image formation process~\cite{Shan_ACMTG_08}.

The accurate motion estimation requirement for video SR has been relaxed
by multidimensional  kernel  regression~\cite{Takeda_TIP09}. Here each pixel is approximated by 3D local series, whose coefficients are estimated by solving a least-square problem, where weights were introduced based on
3D space-time orientation in the pixel neighborhood~\cite{Takeda_TIP09}. However, this
method still assumes the blur kernel that was involved in the LR image
formation model. To do away with the assumptions of blur kernel along with motion and noise level a Bayesian strategy has been developed~\cite{Liu_TPAMI14}, where the
degradation parameters are estimated from the given image sequence.
Motion blur due to relative motion between camera and object can be handled
by considering the least blurred pixels through an EM framework~\cite{Ziyang_CVPR15}.

\subsubsection{Deep Learning Based Approaches}
The motion information among the frames has been explored by BRCN~\cite{Huang_NIPS15}, which modeled long-term contextual information of frames using recurrent neural networks.
Three types of convolutions (feed-forward, recurrent, and conditional) were involved to meet with requirement of spatial dependency, long-term temporal and contextual dependencies. However, iterative motion estimation procedure
increases the computational burden. To reduce the computing load, DECN~\cite{Liao_ICCV15}
introduced non-iterative framework, where different hand-crafted optical flow algorithms are used to generate different SR versions, which were
finally combined using a deep network. Similar technique was employed
in VSRnet~\cite{Kappeler_TCI16}, where motion was compensated in the LR frames by using an optical flow algorithm. The pre-processed LR frames were sent to a pre-trained deep network to generate the final result. Real-time estimation 
of motions among the input LR frames was tried out by VESPCN~\cite{Caballero_CVPR17}, which
is an end-to-end deep network that extracts the optical flow by warping frames using a spatial transformer~\cite{Jaderberg_NIPS15}. Here, the HR frames are produced by another deep network.

Motion estimation and compensation has been practiced in many video SR
approaches. However, Liu et al.~\cite{Liu_ICCV17} have demonstrated that adaptive usage of the motion information in various temporal radius of a temporal adaptive neural network can produce better results. Several inference branches were
used for each temporal radius. The resultant images from these branches
are assembled to produce the final result~\cite{Liu_ICCV17}. The motion compensation
module of VESPCN~\cite{Caballero_CVPR17} was employed in~\cite{Tao_ICCV17}, where sub-pixel motion 
compensation layer was introduced to perform simultaneous motion compensation and resolution enhancement. After the motion compensation, 
 the frames are super-resolved using an encoder-decoder framework with
 skip connections~\cite{Jiao_NIPS16} and ConvLSTM~\cite{Xingjian_NIPS15} module for faster convergence 
 as well as to take care of the sequential nature of video.
Previous end-to-end CNN based video SR methods have focused on explicit motion estimation and compensation to better re-construct  HR  frames.   Unlike  the  previous  works, our approach does not require explicit motion estimation and compensation steps.

\section{Architecture Design}
\label{sec:architecture}

Success of recent approaches has emphasized the importance of network design. Specifically, the most recent image and video SR approaches are built upon two popular image classification networks: residual networks \cite{he2016deep} and densely connected networks \cite{huang2017densely}. These two network designs have also enjoyed success and achieved state-of-the-art performance in other image restoration tasks such as image denoising, dehazing and deblurring. Motivated by the generalization capability of such advances in network designs, our work explores further improvements in network engineering which enable efficient extraction of high-resolution features from low-resolution images. 

\subsection{Proposed Architecture}
The effectiveness of
residual and dense connections has been proved via the significant success in various
vision tasks yet, they cannot be considered as the optimum topology.  For example, too many additions on the same
feature  space  may  impede  the  information  flow  in  ResNet
\cite{huang2017densely}, and there may be the same type of raw features from different layers, which leads to a certain redundancy in DenseNet. Recent works of \cite{chen2017dual,wang2018mixed} partially addressed these issues and demonstrate  improvement in image classification performance. However, very few works have explored optimal network topologies for low-level vision tasks (e.g., image and video SR). We build upon the understanding of dense-topology to design a network for the task of super-resolution that benefits from a mixture of such connections and term it as Mixed-Dense Connection Network (MDCN). 

\begin{figure}[t]
\centering
\includegraphics[scale=0.4]{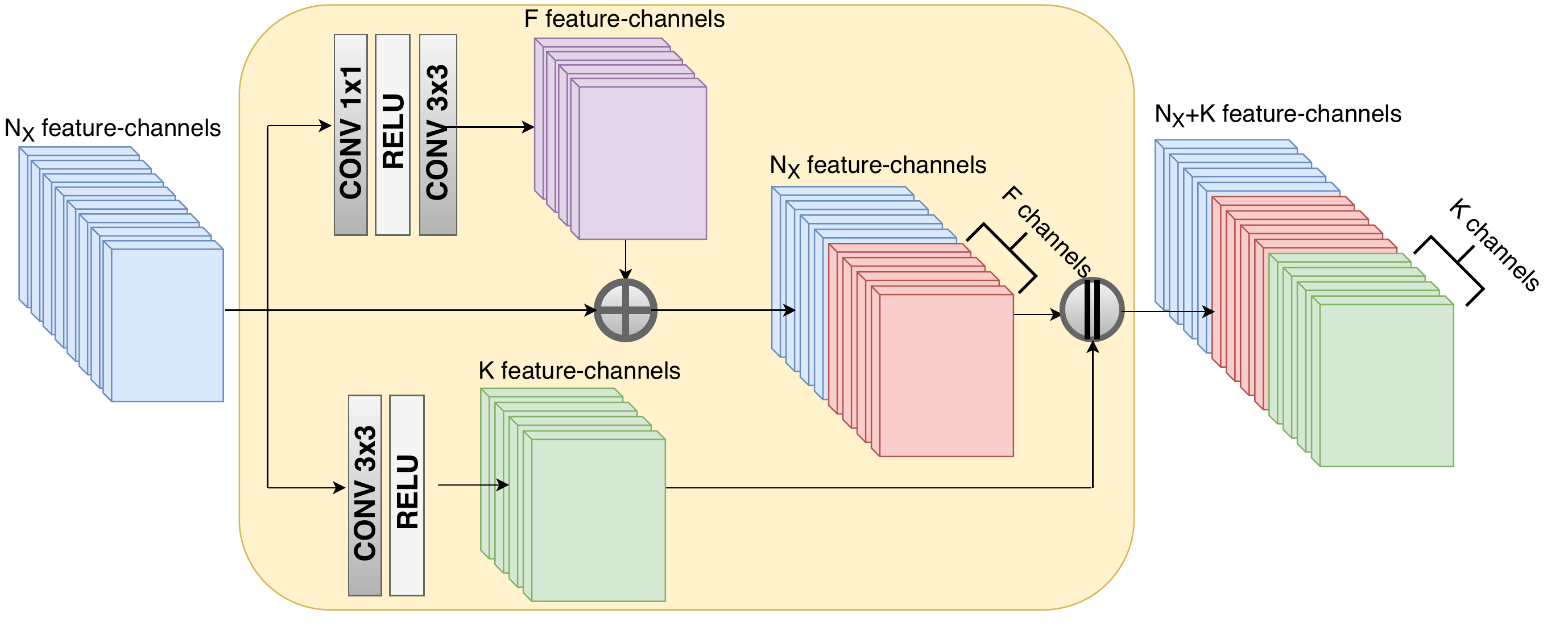} \\
{\small (a) Structure of our {\it Dual Link} unit. The first link selectively modifies $F$ existing feature-maps through addition, while the second link adds $K$ new features through concatenation.}\\
\vspace{3mm}
\includegraphics[scale=0.5]{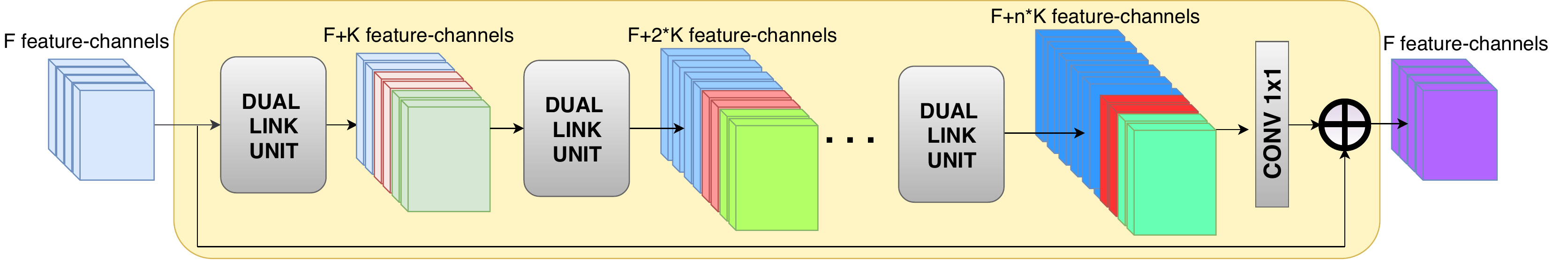} \\ 
{\small (b) Design of our mixed-dense connection block (MDCB). The features channels in red-shade correspond to the channels modified by the additive connection. The features channels in green-shade correspond to the channels added due to concatenation connection.}
\vspace{0mm}
\caption{Structural details of our mixed-dense connection block (MDCB). Each block accepts the darker shade in the feature channels corresponds to deeper and rich features extracted from the initial feature map.}
\label{fig:MDCB}
\vspace{-2mm}
\end{figure}

Our MDCN not only achieves higher accuracy but also enjoys higher parameter efficiency than the state-of-the-art SR approaches. Its strength lies in its building blocks called Mixed-Dense Connection Blocks (MDCBs) which contain a rich set of connections to enable efficient feature-extraction and ease the gradient propagation. Inclusion of addition and
concatenation based connections improves classification accuracy, and is more effective
than going deeper or wider, given an option to increase the capacity of the network. In each MDCB, $n$ Dual Link Units are present. Additive links in the unit grant the benefits of reusing common features with low redundancy, while concatenation links give the network more flexibility in learning new features. Although additive link is flexible with their positions and sizes, each Dual Link Unit performs the additive operation to the last F features of the input. Within each unit, the number of features for additive connections is $F$ and the concatenating connections is $K$. A visual depiction of these connections can be seen in the Fig. \ref{fig:MDCB}(a). This unit adds $K$ new feature maps to the input. The number of features in the input to each MDCB is $F$ as shown in Fig. \ref{fig:MDCB}(b) and after $n$ units, the feature maps contains $F+n*K$ channels. The growth-rate $K$ of concatenation connections was shown to affect the image classification performance positively in \cite{chen2017dual,wang2018mixed}. Further, it was experimentally demonstrated in \cite{zhang2018residual} that deep networks containing many dense blocks stacked together are difficult to train for image restoration and result in poor performance. To handle this, we utilize a gating mechanism to allow larger growth rate by reducing the number of features and hence stabilizes the training of wide network. Each convolution or deconvolution layer is followed by a rectified linear unit (ReLU) for nonlinear mapping, except for the final $1\times1$ layer.

In summary, the advantage of mixed-dense connections manifests in form of significant improvement in propagating error gradients during training. The block has two advantages: Firstly, existing feature channels get modified to learn the residuals: helping in deeper and hierarchical feature extraction. Secondly, feature concatenation with moderate growth-rate promotes new feature exploration. While having these advantages, it also handles the disadvantages: moderate growth rate leads to reduction of redundant features.

\begin{figure}[t]
\centering
\includegraphics[scale=0.52]{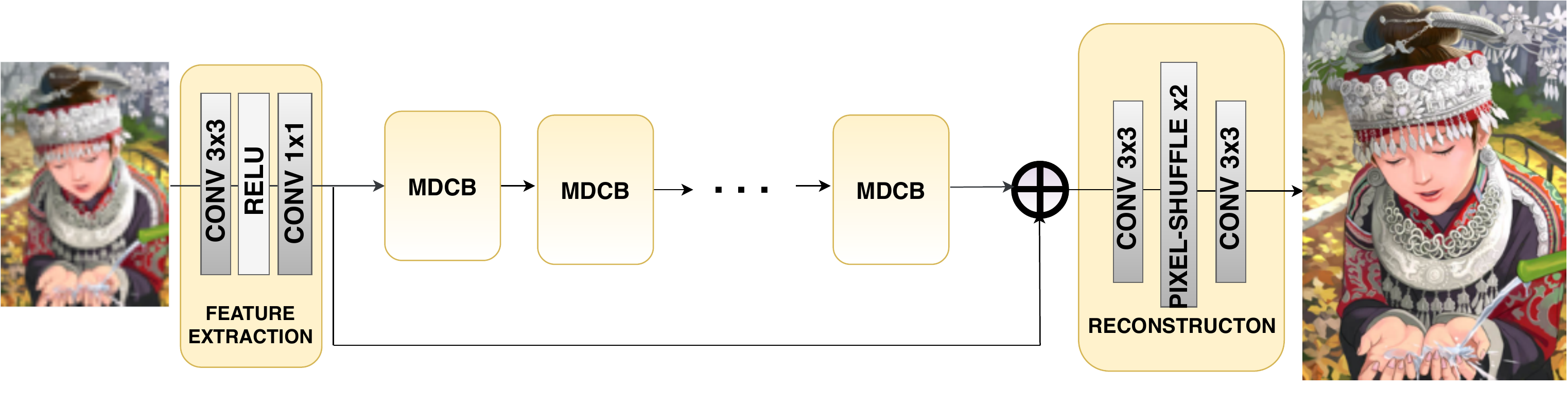}
\vspace{-1mm}
\caption{Architecture of our Mixed Dense Connection Network (MDCN).}
\label{fig:network_MDCN}
\vspace{-4mm}
\end{figure}

Our complete network (MDCN) broadly consists three parts: initial feature extraction module, a series of mixed-dense connection block (MDCBs) and an HR reconstruction module, as shown in Fig.\ref{fig:network_MDCN}. Specifically, the feature extraction module contains two convolutional layers to extract basic feature-maps containing $F$ channels, which are fed to the first MDCB. The HR reconstruction module contains a convolution layer which takes $F$ chanenls features in LR space to $4*F$ channels. The pixel-shuffle layer accepts $4*F$ channels in LR space and returns $F$ channels in HR space, which are passed to the final conv layer to obtain the 3-channel HR image.
\\
\textit{Initial upsampling:} A few approaches interpolate the original LR image to the desired size to form the input to the network. This pre-processing step not only increases computation complexity, but also loses some details of the original LR image. Our network operates on the LR image, extracting deep hierarchical features from it and feeds them to a pixel-shuffle layer to result into the HR image features.
\\
\textit{Filter Size:} A large receptive field is important for utilizing spatial inter-dependencies in the structures present in the LR image. Since receptive-field of the network increases with network depth and/or filter size, depth of the network plays an crucial role in the super-resolution performance, as is evident from the existing state-of-the-art. A few initial works \cite{dong2016image} did utilize larger filter size (larger than $3\times3$) to compensate for smaller depth, but showed only limited super-resolving performance. This can be attributed to the superior speed and parametric efficiency of a $3\times3$ filters above higher-size kernels. A stack of $3\times3$ filters is capable of learning more complex mapping via the non-linearity and the composition of abstraction. Hence, we use $3\times3$ filters for all the layers in our network.
\\
\textit{Differences with \cite{chen2017dual},\cite{wang2018mixed}}: While \cite{chen2017dual} and \cite{wang2018mixed} are proposed for a high-level computer vision task (object recognition), MDCN is suited for image restoration tasks and adopts the concept of blending a variety of dense topological connections at both macro and micro level. Other ncessary design changes include removal of the batch-normalization (BN) layers since they consume the same amount of GPU memory as convolutional layers, increase computational complexity, and hinder performance for image super-resolution. We also remove the pooling layers, since they could discard pixel-level information necessary for super-resolution. To enable higher growth rate, each MDCB contains with a $1\times1$ convolution layer, which fuses the information from the large number of features channels into a fewer feature channels and feeds these to the next MDCB block. Further each MDCB contains a third connection in the form of an additive operation between its input features and the output features.

\subsection{Scale-Recurrent Design}

Most existing SR algorithms treat super-resolution of different scale factors as independent problems without considering and utilizing mutual relationships among different scales in SR. Examples include EDSR~\cite{lim2017enhanced}, which require many scale-specific networks that need to be trained independently to deal with various scales. On the other hand, architectures like VDSR~\cite{kim2016accurate} can handle super-resolution of several scales jointly in the single network. Training the VDSR model with multiple scales boosts the performance and outperforms scale-specific training, implying the redundancy among scale-specific models. Nonetheless, VDSR style architecture takes bicubic interpolated image as the input, which contains less information (as compared to the true LR image) leading to sub-optimal feature extraction; and performs convolutions in HR space, leading to higher computation time and memory requirements. 

\begin{figure}[t]
\centering
\includegraphics[scale=0.52]{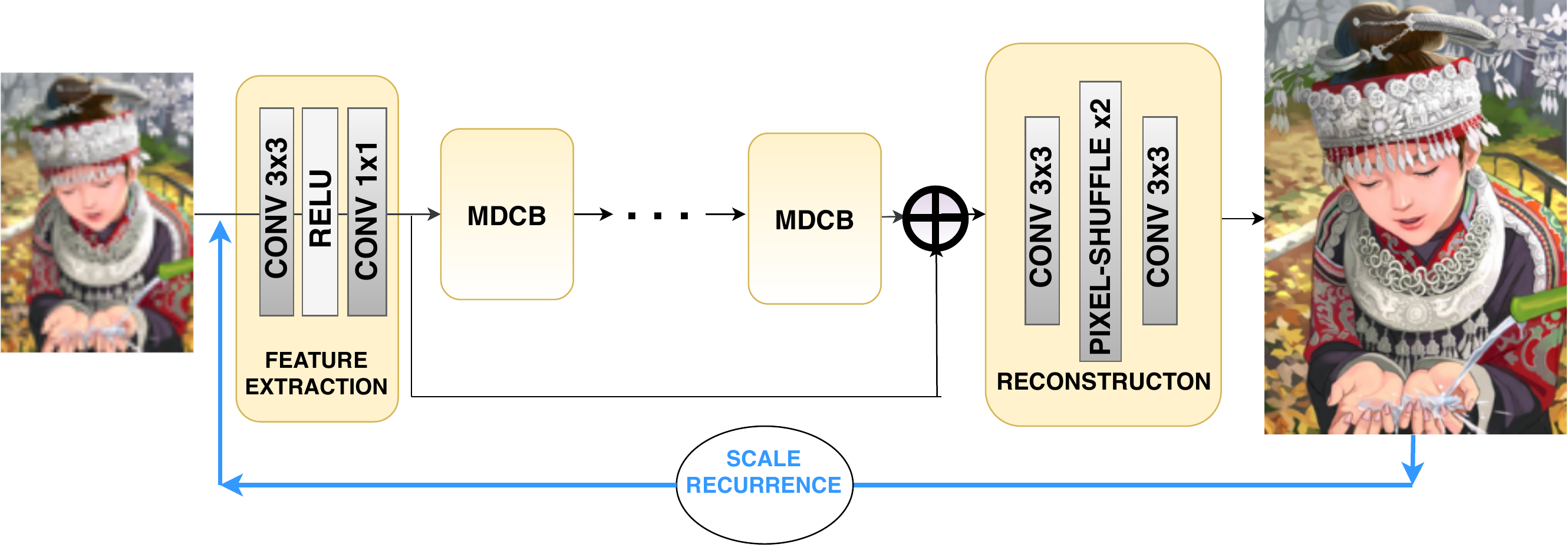}
\vspace{-1mm}
\caption{Scale-Recurrent utilization of our Mixed Dense Connection Network (MDCN). Output of $2\times$ network is fed recursively for $4\times$ and $8\times$ super-resolution.}
\label{fig:network_MDCN_scale_recurrent}
\vspace{-4mm}
\end{figure}

We propose a scale-recurrent training scheme which has the advantages of the approaches mentioned above, while subduing their limitations. Our network’s global design is a multi-scale pyramid which recursively uses the same convolutional filters across the scales, motivated from the fact that a network capable of super-resolving an image by a factor of $2$ can be recursively used to super-resolve the image by a factor $2^s$, $s = 2, 3, \cdots$ Even with the same training data, the recurrent
exploitation of shared weights works in a way similar to using data multiple times to learn parameters, which actually amounts to data augmentation regarding scales. We design the network to reconstruct HR images in intermediate steps by progressively performing a $2\times$ up-sampling of the input from the previous level. Specifically, we first train a network to perform SR by a factor of $2$ and then re-utilize the same weights to take the output of $2\times$ as input and result into a output at resolution $4\times$. This architecture is then fine-tuned to perform $4\times$ SR. We experimentally found that such initialization (training for task of $2\times$ SR) leads to better convergence of the networks for larger scale
factor. Ours is the first approach that efficiently re-utilizes the parameters across scales, which significantly reduces the number of trainable parameters while yielding performance gains for higher scale factors. An illustration of this scheme is shown in Fig. \ref{fig:network_MDCN_scale_recurrent}.

\section{Experimental Results on Single Image SR}
\label{sec:experiments-SISR}
Here, we evaluate our approach on standard datasets for different scale-factors. After discussing different experimental settings, we illustrate
the results of our network, and compare them quantitatively as well as qualitatively with state-of-the-art approaches. The section concludes by
meeting with the perception-distortion trade-off.

\subsection{Settings}
\label{subsec:settings}
We next indicate the experimental settings about datasets, degradation models,  training settings, implementation details, and evaluation metrics. 

\begin{itemize}
\item \textbf{Datasets and degradation model:}
Following~\cite{timofte2017ntire,lim2017enhanced,zhang2018residual,zhang2018learning}, we use 800 training images from DIV2K dataset~\cite{timofte2017ntire} as training set. For testing, we use five standard benchmark datasets: Set5~\cite{bevilacqua2012low}, Set14~\cite{zeyde2012single}, B100~\cite{martin2001database}, Urban100~\cite{huang2015single}, and Manga109~\cite{matsui2017sketch}. We have considered bicubic interpolation for 
generating LR images, corresponding to the HR examples.

\item \textbf{Training settings and implementation details:} 
Network's number of filters are chosen to reduce the number of parameters while not sacrificing the performance super-resolution. We choose a small number of feature-channels as input to each MDCN $F=64$ and a modest growth-rate of $K=36$. Number of MDCBs in hte network is set to $12$ and the number of Dual Link Units within each MDCB is $6$. Data augmentation is performed on the 800 training images, which are randomly rotated by 90$^{\circ}$, 180$^{\circ}$, 270$^{\circ}$ and flipped horizontally. In each training batch, 16 LR color patches with the size of $32\times32$ are extracted and fed as inputs. Our model is trained using ADAM optimizor~\cite{kingma2014adam} with $\beta_{1}=0.9$, $\beta_{2}=0.999$, and $\epsilon=10^{-8}$. The initial leaning rate is set to $10^{-4}$ and then decreases to half every $2\times10^{5}$ iterations of back-propagation. We use PyTorch~\cite{paszke2017automatic} to implement our models with a Titan Xp GPU.

\item \textbf{Evaluation metrics:} 
We have considered PSNR and SSIM~\cite{wang2004image} for comparing our
approach with state-of-the-arts. Note that higher PSNR and SSIM values indicate better quality. Following existing methods, we first converted the image from RGB color space to YCbCr color space, and then calculated the quality assessment metrics using the Y channel (luminance component). This is done by considering the significance of luminance component in visual perception of a scene as compared to the chromatic counterparts. For SR by factor $s$ , we crop $s$ pixels near image boundary before evaluation as in \cite{lim2017enhanced}. Some
of the existing networks such as SRCNN~\cite{dong2016image}, FSRCNN~\cite{dong2016accelerating}, VDSR~\cite{kim2016accurate}, and EDSR~\cite{lim2017enhanced} did not perform $8\times$ super-resolution. To this end, we retrained the existing networks by using author’s code with the recommended parameters.

\item \textbf{Comparisons:}
We considered bicubic interpolation technique and nine deep-learning based approaches for comparisons.
These are SRCNN~\cite{dong2016image}, FSRCNN~\cite{dong2016accelerating}, VDSR~\cite{kim2016accurate}, LapSRN~\cite{lai2017deep}, MemNet~\cite{tai2017memnet}, EDSR~\cite{lim2017enhanced}, SRMDNF~\cite{zhang2018learning}, D-DBPN~\cite{haris2018deep} and RDN~\cite{zhang2018residual}. For perceptual and objective super-resolution, we have performed qualitative comparisons with SRResNet,SRGAN~\cite{ledig2017photo} and ENet-E,Enet-PAT~\cite{sajjadi2017enhancenet} in subsection 5.3.
We have also encompassed the geometric self-ensemble strategy in testing our network for improvement, as has been done in~\cite{lim2017enhanced,zhang2018residual}.
\end{itemize}

\subsection{Results on Standard Benchmarks}

\begin{table}[th!]
\scriptsize
\center
\begin{center}
\caption{Quantitative results with bi-cubic degradation model. Best results are \textbf{highlighted}.}
\label{tab:quant}
\begin{tabular}{|l|c|c|c|c|c|c|c|c|c|c|c|c|}
\hline
 
\multirow{2}{*}{Method} & \multirow{2}{*}{Training Set} & \multirow{2}{*}{Scale} &  \multicolumn{2}{c|}{Set5} &  \multicolumn{2}{c|}{Set14} &  \multicolumn{2}{c|}{B100} &  \multicolumn{2}{c|}{Urban100} &  \multicolumn{2}{c|}{Manga109}  
\\
\cline{4-13}
& & & PSNR & SSIM & PSNR & SSIM & PSNR & SSIM & PSNR & SSIM & PSNR & SSIM 
\\
\hline
\hline
Bicubic & - & $\times$2
& 33.66
 & 0.9299
  & 30.24
   & 0.8688
    & 29.56
     & 0.8431
      & 26.88
       & 0.8403
        & 30.80
         & 0.9339
                  
\\
SRCNN~\cite{dong2016image} & 91+ImageNet & $\times$2 
& 36.66
 & 0.9542
  & 32.45
   & 0.9067
    & 31.36
     & 0.8879
      & 29.50
       & 0.8946
        & 35.60
         & 0.9663
                   
\\
FSRCNN~\cite{dong2016accelerating} & 291 & $\times$2 
& 37.05
 & 0.9560
  & 32.66
   & 0.9090
    & 31.53
     & 0.8920
      & 29.88
       & 0.9020
        & 36.67
         & 0.9710
                   
\\
VDSR~\cite{kim2016accurate} & 291 & $\times$2 
& 37.53
 & 0.9590
  & 33.05
   & 0.9130
    & 31.90
     & 0.8960
      & 30.77
       & 0.9140
        & 37.22
         & 0.9750
                   
\\
LapSRN~\cite{lai2017deep} & 291 & $\times$2 
& 37.52
 & 0.9591
  & 33.08
   & 0.9130
    & 31.08
     & 0.8950
      & 30.41
       & 0.9101
        & 37.27
         & 0.9740
                   
\\
MemNet~\cite{tai2017memnet} & 291 & $\times$2 
& 37.78
 & 0.9597
  & 33.28
   & 0.9142
    & 32.08
     & 0.8978
      & 31.31
       & 0.9195
        & 37.72
         & 0.9740
                   
\\
EDSR~\cite{lim2017enhanced} & DIV2K & $\times$2 
& 38.11
 & 0.9602
  & 33.92
   & 0.9195
    & 32.32
     & 0.9013
      & 32.93
       & 0.9351
        & 39.10
         & 0.9773
                   
\\
SRMDNF~\cite{zhang2018learning} & DIV2K+WED  & $\times$2 
& 37.79
 & 0.9601
  & 33.32
   & 0.9159
    & 32.05
     & 0.8985
      & 31.33
       & 0.9204
        & 38.07
         & 0.9761
                   
\\
D-DBPN~\cite{haris2018deep} & DIV2K+Flickr & $\times$2 
& 38.09
 & 0.9600
  & 33.85
   & 0.9190
    & 32.27
     & 0.9000
      & 32.55
       & 0.9324
        & 38.89
         & 0.9775        
\\
RDN~\cite{zhang2018residual} & DIV2K  & $\times$2 
& 38.24
 & 0.9614
  & 34.01
   & 0.9212
    & 32.34
     & 0.9017
      & 32.89
       & 0.9353
        & 39.18
         & 0.9780
         
\\

MDCN (ours) & DIV2K  & $\times$2 
& 38.24
 & 0.9613
  & 33.93
   & 0.9207
    & 32.34
     & 0.9017
      & 32.83
       & 0.9353
        & 39.09
         & 0.9777

\\
MDCN+ (ours) & DIV2K  & $\times$2 
& \textbf{38.30}
 & \textbf{0.9616}
  & \textbf{34.05}
   & \textbf{0.9217}
    & \textbf{32.39}
     & \textbf{0.9023}
      & \textbf{33.05}
       & \textbf{0.9369}
        & \textbf{39.32}
         & \textbf{0.9783}

\\
\hline
\hline
Bicubic & - & $\times$3 
& 30.39
 & 0.8682
  & 27.55
   & 0.7742
    & 27.21
     & 0.7385
      & 24.46
       & 0.7349
        & 26.95
         & 0.8556
                  
\\
SRCNN~\cite{dong2016image} & 91+ImageNet & $\times$3
& 32.75
 & 0.9090
  & 29.30
   & 0.8215
    & 28.41
     & 0.7863
      & 26.24
       & 0.7989
        & 30.48
         & 0.9117
                    
\\
FSRCNN~\cite{dong2016accelerating} & 291 & $\times$3 
& 33.18
 & 0.9140
  & 29.37
   & 0.8240
    & 28.53
     & 0.7910
      & 26.43
       & 0.8080
        & 31.10
         & 0.9210
                   
\\
VDSR~\cite{kim2016accurate} & 291 & $\times$3 
& 33.67
 & 0.9210
  & 29.78
   & 0.8320
    & 28.83
     & 0.7990
      & 27.14
       & 0.8290
        & 32.01
         & 0.9340
                   
\\
LapSRN~\cite{lai2017deep} & 291 & $\times$3 
& 33.82
 & 0.9227
  & 29.87
   & 0.8320
    & 28.82
     & 0.7980
      & 27.07
       & 0.8280
        & 32.21
         & 0.9350
                   
\\
MemNet~\cite{tai2017memnet} & 291 & $\times$3 
& 34.09
 & 0.9248
  & 30.00
   & 0.8350
    & 28.96
     & 0.8001
      & 27.56
       & 0.8376
        & 32.51
         & 0.9369
                   
\\
EDSR~\cite{lim2017enhanced} & DIV2K & $\times$3 
& 34.65
 & 0.9280
  & 30.52
   & 0.8462
    & 29.25
     & 0.8093
      & 28.80
       & 0.8653
        & 34.17
         & 0.9476
                   
\\
SRMDNF~\cite{zhang2018learning} & DIV2K+WED & $\times$3 
& 34.12
 & 0.9254
  & 30.04
   & 0.8382
    & 28.97
     & 0.8025
      & 27.57
       & 0.8398
        & 33.00
         & 0.9403
                   
\\
RDN~\cite{zhang2018residual} & DIV2K & $\times$3 
& 34.71
 & 0.9296
  & 30.57
   & 0.8468
    & 29.26
     & 0.8093
      & 28.80
       & 0.8653
        & 34.13
         & 0.9484
         
\\

MDCN (ours) & DIV2K & $\times$3 
& 34.71
 & 0.9295
  & 30.58
   & 0.8473
    & 29.28
     & 0.8096
      & 28.75
       & 0.8656
        & 34.18
         & 0.9485

\\
MDCN+ (ours) & DIV2K & $\times$3 
& \textbf{34.82}
 & \textbf{0.9303}
  & \textbf{30.68}
   & \textbf{0.8486}
    & \textbf{29.34}
     & \textbf{0.8106}
      & \textbf{28.96}
       & \textbf{0.8686}
        & \textbf{34.50}
         & \textbf{0.9500}
         
\\
\hline
\hline
Bicubic & - & $\times$4 
& 28.42
 & 0.8104
  & 26.00
   & 0.7027
    & 25.96
     & 0.6675
      & 23.14
       & 0.6577
        & 24.89
         & 0.7866
                  
\\
SRCNN~\cite{dong2016image} & 91+ImageNet & $\times$4 
& 30.48
 & 0.8628
  & 27.50
   & 0.7513
    & 26.90
     & 0.7101
      & 24.52
       & 0.7221
        & 27.58
         & 0.8555
                   
\\
FSRCNN~\cite{dong2016accelerating} & 291 & $\times$4 
& 30.72
 & 0.8660
  & 27.61
   & 0.7550
    & 26.98
     & 0.7150
      & 24.62
       & 0.7280
        & 27.90
         & 0.8610
                   
\\
VDSR~\cite{kim2016accurate} & 291 & $\times$4 
& 31.35
 & 0.8830
  & 28.02
   & 0.7680
    & 27.29
     & 0.0726
      & 25.18
       & 0.7540
        & 28.83
         & 0.8870
                   
\\
LapSRN~\cite{lai2017deep} & 291 & $\times$4 
& 31.54
 & 0.8850
  & 28.19
   & 0.7720
    & 27.32
     & 0.7270
      & 25.21
       & 0.7560
        & 29.09
         & 0.8900
                   
\\
MemNet~\cite{tai2017memnet} & 291 & $\times$4 
& 31.74
 & 0.8893
  & 28.26
   & 0.7723
    & 27.40
     & 0.7281
      & 25.50
       & 0.7630
        & 29.42
         & 0.8942
                   
\\
EDSR~\cite{lim2017enhanced} & DIV2K & $\times$4 
& 32.46
 & 0.8968
  & 28.80
   & 0.7876
    & 27.71
     & 0.7420
      & 26.64
       & 0.8033
        & 31.02
         & 0.9148                 
\\
SRMDNF~\cite{zhang2018learning} & DIV2K+WED & $\times$4 
& 31.96
 & 0.8925
  & 28.35
   & 0.7787
    & 27.49
     & 0.7337
      & 25.68
       & 0.7731
        & 30.09
         & 0.9024
                   
\\
D-DBPN~\cite{haris2018deep} & DIV2K+Flickr & $\times$4 
& 32.47
 & 0.8980
  & 28.82
   & 0.7860
    & 27.72
     & 0.7400
      & 26.38
       & 0.7946
        & 30.91
         & 0.9137
         
\\
RDN~\cite{zhang2018residual} & DIV2K & $\times$4 
& 32.47
 & 0.8990
  & 28.81
   & 0.7871
    & 27.72
     & 0.7419
      & 26.61
       & 0.8028
        & 31.00
         & 0.9151
         
\\

MDCN (ours) & DIV2K & $\times$4 
& 32.59
 & 0.8994
  & 28.84
   & 0.7877
    & 27.73
     & 0.7416
      & 26.62
       & 0.8030
        & 31.03
         & 0.9160

\\
MDCN+ (ours) & DIV2K & $\times$4 
& \textbf{32.66}
 & \textbf{0.9003}
  & \textbf{28.92}
   & \textbf{0.7893}
    & \textbf{27.79}
     & \textbf{0.7428}
      & \textbf{26.785}
       & \textbf{0.8065}
        & \textbf{31.34}
         & \textbf{0.9184}
              
\\
\hline
\hline
Bicubic & - & $\times$8 
& 24.40
 & 0.6580
  & 23.10
   & 0.5660
    & 23.67
     & 0.5480
      & 20.74
       & 0.5160
        & 21.47
         & 0.6500
                 
\\
SRCNN~\cite{dong2016image} & 91+ImageNet & $\times$8 
& 25.33
 & 0.6900
  & 23.76
   & 0.5910
    & 24.13
     & 0.5660
      & 21.29
       & 0.5440
        & 22.46
         & 0.6950
                   
\\
FSRCNN~\cite{dong2016accelerating} & 291 & $\times$8 
& 20.13
 & 0.5520
  & 19.75
   & 0.4820
    & 24.21
     & 0.5680
      & 21.32
       & 0.5380
        & 22.39
         & 0.6730
                   
\\
SCN~\cite{wang2015deep} & 91 & $\times$8 
& 25.59
 & 0.7071
  & 24.02
   & 0.6028
    & 24.30
     & 0.5698
      & 21.52
       & 0.5571
        & 22.68
         & 0.6963

\\
VDSR~\cite{kim2016accurate} & 291 & $\times$8 
& 25.93
 & 0.7240
  & 24.26
   & 0.6140
    & 24.49
     & 0.5830
      & 21.70
       & 0.5710
        & 23.16
         & 0.7250
                   
\\   
LapSRN~\cite{lai2017deep} & 291 & $\times$8 
& 26.15
 & 0.7380
  & 24.35
   & 0.6200
    & 24.54
     & 0.5860
      & 21.81
       & 0.5810
        & 23.39
         & 0.7350
                   
\\
MemNet~\cite{tai2017memnet} & 291 & $\times$8 
& 26.16
 & 0.7414
  & 24.38
   & 0.6199
    & 24.58
     & 0.5842
      & 21.89
       & 0.5825
        & 23.56
         & 0.7387

\\
MSLapSRN~\cite{MSLapSRN} & 291 & $\times$8 
& 26.34
 & 0.7558
  & 24.57
   & 0.6273
    & 24.65
     & 0.5895
      & 22.06
       & 0.5963
        & 23.90
         & 0.7564
                   
\\
EDSR~\cite{lim2017enhanced} & DIV2K & $\times$8 
& 26.96
 & 0.7762
  & 24.91
   & 0.6420
    & 24.81
     & 0.5985
      & 22.51
       & 0.6221
        & 24.69
         & 0.7841
                   
\\
D-DBPN~\cite{haris2018deep} & DIV2K+Flickr & $\times$8 
& 27.21
 & 0.7840
  & 25.13
   & 0.6480
    & 24.88
     & 0.6010
      & 22.73
       & 0.6312
        & 25.14
         & 0.7987
         
\\

MDCN (ours) & DIV2K & $\times$8 
& 27.32
 & 0.7867
  & 25.13
   & 0.6472
    & 24.93
     & 0.6023
      & 22.84
       & 0.6370
        & 25.00
         & 0.7956

\\
MDCN+ (ours) & DIV2K & $\times$8 
& \textbf{27.39}
 & \textbf{0.7895}
  & \textbf{25.25}
   & \textbf{0.6499}
    & \textbf{24.99}
     & \textbf{0.6039}
      & \textbf{23.02}
       & \textbf{0.6422}
        & \textbf{25.26}
         & \textbf{0.8009}
           
\\
\hline             
\end{tabular}
\end{center}
\end{table}

\begin{figure}[th!]
	\newlength\fsdurthree
	\setlength{\fsdurthree}{-1.5mm}
	\scriptsize
	\centering
	\begin{tabular}{ccc}
	
		\begin{adjustbox}{valign=t}
		\begin{tabular}{c}
		\tiny
				\includegraphics[width=0.309\textwidth]{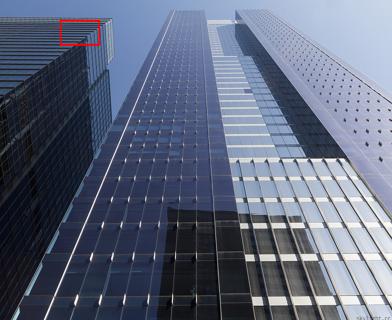} 
				 \\
				 Urban 100 (3X)

		\end{tabular}
		\end{adjustbox}\\
		
		\begin{adjustbox}{valign=t}
		\tiny
			\begin{tabular}{cc}
				\includegraphics[width=\widthscalefive \textwidth]{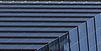} \hspace{\fsdurthree} &
				\includegraphics[width=\widthscalefive \textwidth]{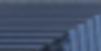}  \hspace{\fsdurthree} 
				\\
HR  \hspace{\fsdurthree} & Bicubic \hspace{\fsdurthree}
\\

				\includegraphics[width=\widthscalefive \textwidth]{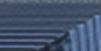} \hspace{\fsdurthree} &
				\includegraphics[width=\widthscalefive \textwidth]{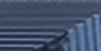}  \hspace{\fsdurthree} \\
SRCNN~\cite{dong2016image} \hspace{\fsdurthree} &
FSRCNN~\cite{dong2016accelerating} \hspace{\fsdurthree} \\

				\includegraphics[width=\widthscalefive \textwidth]{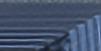} &
						\includegraphics[width=\widthscalefive \textwidth]{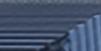} \hspace{\fsdurthree}
					
						 \\
VDSR~\cite{kim2016accurate}\hspace{\fsdurthree} & IRCNN~\cite{IRCNN} \hspace{\fsdurthree}  \hspace{\fsdurthree}\\

 				\includegraphics[width=\widthscalefive \textwidth]{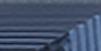} \hspace{\fsdurthree} &
				\includegraphics[width=\widthscalefive \textwidth]{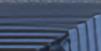}  \hspace{\fsdurthree}
				\\
SPMSR~\cite{SPMSR} \hspace{\fsdurthree} &SRMDNF~\cite{zhang2018learning} \hspace{\fsdurthree}\\

				\includegraphics[width=\widthscalefive \textwidth]{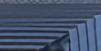} \hspace{\fsdurthree} &
				\includegraphics[width=\widthscalefive \textwidth]{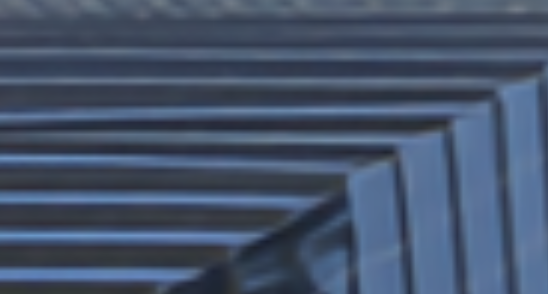} 
				\\
RDN~\cite{zhang2018residual} \hspace{\fsdurthree} &MDCN \hspace{\fsdurthree}
\\

			\end{tabular}
		\end{adjustbox}
		\vspace{-3mm}
	\end{tabular}
	\caption{
		Visual comparison for $3\times$ SR with bi-cubic degradation on Urban100 dataset. }
\label{fig:result_3x_Urban100}
\vspace{-5mm}
\end{figure}
\begin{figure}[ht!]
	\setlength{\fsdurthree}{-1.5mm}
	\scriptsize
	\centering
	\begin{tabular}{ccc}

		\begin{adjustbox}{valign=t}
		\begin{tabular}{c}
		\tiny
			
				\includegraphics[width=0.309\textwidth]{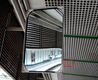} \\
				Urban 100 (4X)

		\end{tabular}
		\end{adjustbox}\\		
		
		\begin{adjustbox}{valign=t}
		\tiny
			\begin{tabular}{cc}
				\includegraphics[width=\widthscalefive \textwidth]{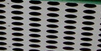} \hspace{\fsdurthree} &
				\includegraphics[width=\widthscalefive \textwidth]{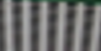} \hspace{\fsdurthree}
				 \hspace{\fsdurthree} 
				\\
HR  \hspace{\fsdurthree} & Bicubic \hspace{\fsdurthree} 
\\

				\includegraphics[width=\widthscalefive \textwidth]{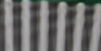} \hspace{\fsdurthree} &
				\includegraphics[width=\widthscalefive \textwidth]{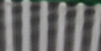} \hspace{\fsdurthree}
				\\
SRCNN~\cite{dong2016image} \hspace{\fsdurthree} &
FSRCNN~\cite{dong2016accelerating} \hspace{\fsdurthree}  \\				

				\includegraphics[width=\widthscalefive \textwidth]{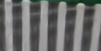} &
						\includegraphics[width=\widthscalefive \textwidth]{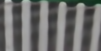} \hspace{\fsdurthree}
						 \\
VDSR~\cite{kim2016accurate}\hspace{\fsdurthree} & LapSRN~\cite{lai2017deep} \hspace{\fsdurthree} \\

 				\includegraphics[width=\widthscalefive \textwidth]{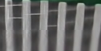} \hspace{\fsdurthree} &
				\includegraphics[width=\widthscalefive \textwidth]{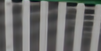} \hspace{\fsdurthree}
				\\
MemNet~\cite{tai2017memnet} \hspace{\fsdurthree} &EDSR~\cite{lim2017enhanced} \hspace{\fsdurthree} \\

				\includegraphics[width=\widthscalefive \textwidth]{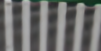} \hspace{\fsdurthree} &
				\includegraphics[width=\widthscalefive \textwidth]{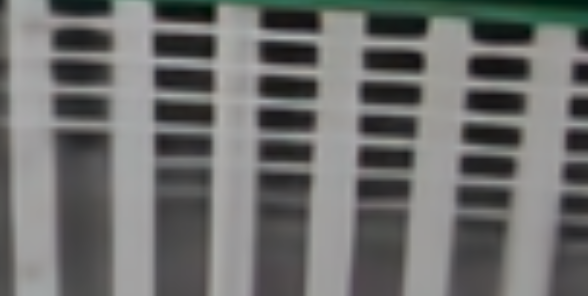} 
	
				\\
SRMDNF~\cite{zhang2018learning} \hspace{\fsdurthree} &MDCN \hspace{\fsdurthree}
\\

			\end{tabular}
		\end{adjustbox}
		\vspace{-3mm}
	\end{tabular}
	\caption{
		Visual comparison for $4\times$ SR with bi-cubic degradation on Urban100 and Manga109 datasets. }
\label{fig:result_4x_Urban100_Manga109}
\vspace{-5mm}
\end{figure}

The quantitative comparisons on the test sets (Set5, Set14, B100, Urban100, \& Manga109) in terms of average PSNR \& SSIM are given in Table~\ref{tab:quant} for scale factors $\times$2, $\times$3, $\times$4, and $\times$8. The results of our network and geometric self-ensemble strategy are shown in MDCN \& MDCN+ rows, respectively. One can note that our MDCN strategy is able to out-perform most of the approaches for lower scale factors 2 \& 3. Whereas,
for higher factors such as 4 \& 8, proposed MDCN+ is able to surpass all the state-of-the-art approaches. Observe that even without the geometric self-ensemble strategy, our MDCN is able to out-perform the existing approaches for higher scale factors.

 However, for the scaling factor ×2, the method of RDN shows close performance  to our network, which can be attributed to their sheer number of parameters ($120\%$ more than our network). When the scaling factor becomes larger (e.g.,$\times3$ and $\times4$), RDN does not hold the similar advantage over our network. In terms of parametric efficiency, we outperform all the methods with a large margin. Parametrically, MDSR is closest to ours, but leads to a lower performance. The parametric efficiency of MDSR over existing approaches like EDSR,RDN and DDBPN can be attributed to their training configuration. First, MDSR utilizes multi-scale inputs as VDSR does \cite{kim2016accurate}, which enables them to estimate and fuse different scale features. Second, MDSR uses larger input patch size (48 against 32) for training. As most images in Urban100 contain self-similar structures, larger input patch size for training allows a very deep network to grasp more information by using large receptive field better. However, in our MDCN, we do not use multi-scale information or larger patch size.
Moreover, our MDCN+ can achieve further improvement with the geometric self-ensemble technique. This is due to our effective combination of residual and dense connections through DBCN block with scale-recurrent strategy.

\begin{figure}[ht!]
	\setlength{\fsdurthree}{-1.5mm}
	\scriptsize
	\centering
	\begin{tabular}{ccc}
		
		\begin{adjustbox}{valign=t}
		\begin{tabular}{c}
		\tiny
				\includegraphics[width=0.309\textwidth]{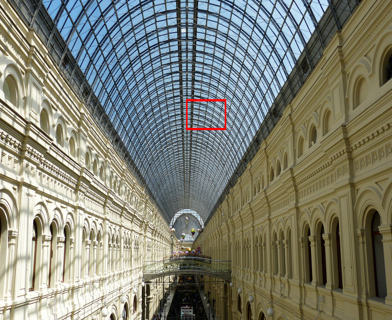} \\
				
				Urban 100 (8X)
				
		\end{tabular}
		\end{adjustbox}\\		
		
		\begin{adjustbox}{valign=t}
		\tiny
			\begin{tabular}{cc}
				\includegraphics[width=\widthscalefive \textwidth]{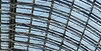} \hspace{\fsdurthree} &
				\includegraphics[width=\widthscalefive \textwidth]{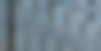} \hspace{\fsdurthree}
				\\
HR  \hspace{\fsdurthree} & Bicubic \hspace{\fsdurthree}
\\
		
				\includegraphics[width=\widthscalefive \textwidth]{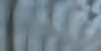} \hspace{\fsdurthree} &
				\includegraphics[width=\widthscalefive \textwidth]{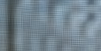} \hspace{\fsdurthree} 
				\\
SRCNN~\cite{dong2016image} \hspace{\fsdurthree} &
SCN~\cite{wang2015deep} \hspace{\fsdurthree} \\

				\includegraphics[width=\widthscalefive \textwidth]{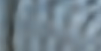} &
						\includegraphics[width=\widthscalefive \textwidth]{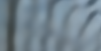} \hspace{\fsdurthree}
						 \\
VDSR~\cite{kim2016accurate}\hspace{\fsdurthree} & LapSRN~\cite{lai2017deep} \hspace{\fsdurthree} \\

 				\includegraphics[width=\widthscalefive \textwidth]{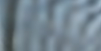} \hspace{\fsdurthree} &
				\includegraphics[width=\widthscalefive \textwidth]{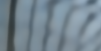} \hspace{\fsdurthree} 
				\\
MemNet~\cite{tai2017memnet} \hspace{\fsdurthree} &MSLapSRN~\cite{zhang2018learning} \hspace{\fsdurthree}\\

				\\
				\includegraphics[width=\widthscalefive \textwidth]{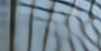} \hspace{\fsdurthree} &
				\includegraphics[width=\widthscalefive \textwidth]{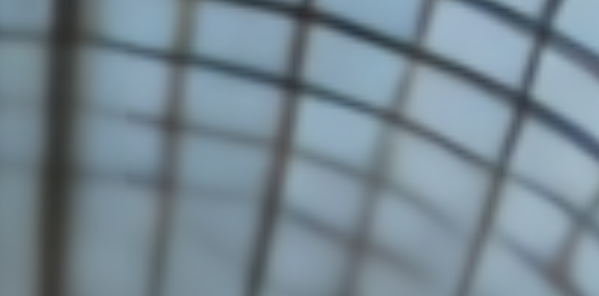} 
				\\
EDSR~\cite{lim2017enhanced} \hspace{\fsdurthree} &MDCN \hspace{\fsdurthree} \hspace{\fsdurthree}
\\

			\end{tabular}
		\end{adjustbox}
		\vspace{-3mm}
	\end{tabular}
	\caption{
		Visual comparison for $8\times$ SR with bi-cubic degradation on Urban100 \& Manga109 datasets.	}
\label{fig:result_8x_Urban100}
\end{figure}

The improvement of results can be observed visually in Figs.~\ref{fig:result_3x_Urban100}, \ref{fig:result_4x_Urban100_Manga109} \& \ref{fig:result_8x_Urban100} for scale factors 3, 4 and 8, respectively. 
In Fig.~\ref{fig:result_3x_Urban100} (img$\_$033), one can observe that most of the approaches produce smoother results as compared to ours. Whereas, for img$\_$037, the approaches yield results, where some of the linear structures are disorientated. However, our MDCN is able to produce result with appropriate line structure. On the other hand, most of the existing methods fail to reproduce the line structures of the roof of the railway station in img$\_$098, whereas our strategy helps in bringing back those details in the result.

For scale factor $4$, the size of the input image itself is too small to 
provide different details to the network. Hence, the existing approaches
fail to maintain different structures faithfully. For example, in img$\_004$ of Fig.~\ref{fig:result_4x_Urban100_Manga109}, the elliptical shape structure is completely demolished and other regions appear to be parallel
structured vertical lines. However, our MDCN is able to maintain somewhat similar structure to the ground-truth. The img$\_092$ consists of lines
mainly in two directions. However, the existing models are not able to 
maintain the line orientation as per with the ground truth. Moreover, 
some of the approaches creates blocky artifacts due to generation of 
some unwanted lines. In contrast, our MDCN is able to maintain the line orientations appropriately. Blurring artifacts are quite evident for such 
larger scale factors, as can be observed in third scene of Fig.~\ref{fig:result_4x_Urban100_Manga109} for the results of most of the existing approaches. The result of our model has less affected by blur.

Such blurring artifact increases with larger scale factor such as 8.
This is because for very high scale factor, the resolution of the input image is quite low with minimum details. Thus,
super-resolving these LR images by a factor of $8$ is very challenging.
As a consequence, most of the approaches suffer from over-smoothing effect, as can be seen in Fig.~\ref{fig:result_8x_Urban100}.
Among the existing methods, EDSR~\cite{lim2017enhanced} can bring out some texture details but the orientations of those are not appropriate. On contrary, our MDCN
is able to bring out texture details with appropriate orientation.
This is due to our elegant scale-recurrent framework, which enables us
to carry-over the improvements for lower scale factors via weight transfer
strategy.

\vspace{0.5cm}

\section{Video Super-Resolution}
\label{sec:videoSR}

We extend our MDCN for video SR through a multi-image restoration approach. 
Although our trained SISR network can be directly utilized for the video SR task by processing each frame of the video individually, access to multiple neighboring LR frames can fundamentally reduce the ill-posedness of the SR task and potentially yield higher reconstruction quality.

We experimentally verified this fact by training a modified version of our network that accepts 5 consecutive LR frames (includes the frame to super-resolve and its 4 neighbors) as input. The frames are concatenated along the channel dimension before being fed to the first layer our network, which is modified to accept 15 channel inputs. This is referred to as early fusion technique in video SR literature.

The source for training data for this task was the Vimeo Super-Resolution (VSR) dataset \cite{xue2017video} that contains $64612$ training samples. Each sample in the dataset contains seven consecutive frames with $448\times 256$ resolution. We addressed the task of $\times4$ SR, where our network can reap the joint benefits of MDCBs and scale-recurrence. We used batch size of $8$, where each sample contained 5 LR frames of $32\times32$ resolution and 1 HR frame of size $128\times128$. Our network was trained using Adam optimizer with a learning rate of $10^{-4}$ for $4\times10^4$ iterations.

\subsection{Experimental results on video SR}
For evaluation, we have considered standard Vid4 dataset, which consists of 4 videos. The results are compared with state-of-the-art video SR approaches such as VSRnet~\cite{Kappeler_TCI16}, Bayesian~\cite{Liu_TPAMI14}, ESPCN~\cite{ESPCN}, VESPCN~\cite{Caballero_CVPR17}, and Temp\_robust~\cite{Robust_temp}. The quality assessment metrics are computed using the results provided by the respective authors of the approaches. The metrics for each of the approaches have been computed by considering 30 frames and by removing 8 boundary pixels from each of them.

\begin{figure*}[!ht]
	\centering
		\begin{tabular}{c@{ }  c@{ } c@{ }  c@{ } c@{ } c@{ } }	
		\includegraphics[trim = 0 0 0 0, clip, scale=0.5]{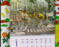} &
		\includegraphics[trim = 0 0 0 0, clip, scale = 1.30]{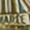} &
		\includegraphics[trim = 0 0 0 0, clip, scale = 0.20]{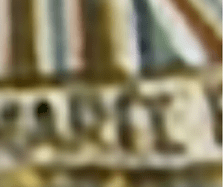} &
		\includegraphics[trim = 0 0 0 0, clip, scale = 0.20]{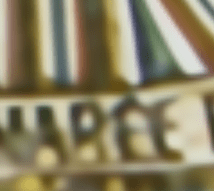} &
				\hspace{-0cm}\includegraphics[trim = 0 0 0 0, clip, scale = 0.21]{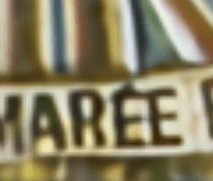} &
		\includegraphics[trim = 0 0 0 0, clip, scale = 0.165]{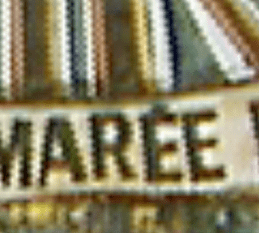} \\
		\includegraphics[trim = 0 0 0 0, clip, scale=0.50]{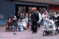} &
		\includegraphics[trim = 0 0 0 0, clip, scale = 1.41]{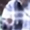} &
		\includegraphics[trim = 0 0 0 0, clip, scale = 0.20]{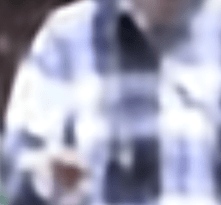} &
		\includegraphics[trim = 0 0 0 0, clip, scale = 0.20]{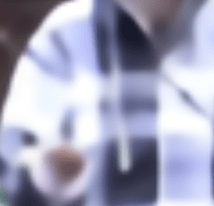} &
		\includegraphics[trim = 0 0 0 0, clip, scale = 0.19]{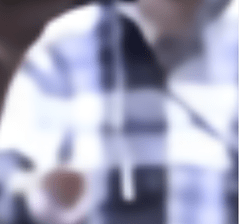} &		
		\includegraphics[trim = 0 0 0 0, clip, scale = 0.19]{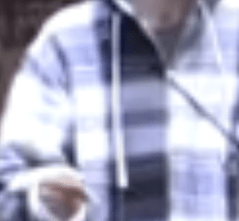}\\		
		(a) & (b) & (c) & (d) & (e) & (f) \\
		\end{tabular}
		\vspace{-0.5em}	
		\caption{\label{fig:video} Comparisons for video SR (x4) on vid4 dataset: (a) LR frame, (b) VESPCN, (c) Temp\_robust, (d) Our model:single-frame based, (d) Our model:multiple-frame based, and (f) ground truth. First row represents the results of a frame of $calendar$ video, and second row depicts the results of a frame of $walk$.}
\end{figure*}

\vspace{0.5cm}
\begin{table}[htbp]
\centering
\begin{center}
\begin{tabular}{|c|c|c|c|c|c|c|}
\hline
Metric & Bicubic & Bayesian & VSRnet & ESPCN & VESPCN  & Ours\\
\hline
PSNR & 25.38 & 25.64 & 26.64 & 26.97 & 27.25 &  {\bf 27.55} \\
SSIM & 0.7613 & 0.8000 & 0.8238 & 0.8364 & {\bf 0.8400} &  0.8332\\
\hline
\end{tabular}
\end{center}
\vspace{-2mm}
\caption{Comparison of video SR results on the standard Vid4 dataset for scale factor 3. Best results are {\bf highlighted}.}
\label{tab:videox3}
\vspace{-1mm}
\end{table}
We have provided average PSNR and SSIM scores of different approaches along with our SISR model in Table~\ref{tab:videox3} for scale factor 3. The comparison shows that our SISR network is able to out-perform the competing videoSR approaches in terms of PSNR for scale factor 3. 
Next, we evaluate our models on the challenging task of $\times4$ video SR. Average PSNR and SSIM score for each video for scale factor 4 on the same dataset are kept in Table~\ref{tab:videox4}. One can observe that for scale factor 4, our MDCN is able to out-perform most of the video SR approaches and produce comparable results to Temp\_robust~\cite{Robust_temp}. This improvement can also be observed visually in Fig.~\ref{fig:video}(d) for
scale factor 4.  

\begin{table}[htbp]
\centering
\begin{center}
\begin{tabular}{|c|c|c|c|c|c|c|c|c|}
\hline
Video  & Metric & Bi-cubic & VSRnet & ESPCN & VESPCN & Temp\_robust & Ours  & Ours(multi-frame)
\\
\hline \hline
Calendar & PSNR& 20.51 & 21.27 & 21.69 & 21.92 & 22.14 & { 22.38} & {\bf 23.36}\\
 & SSIM & 0.5622 & 0.6390 & 0.6736 & 0.6858 & 0.7052 & { 0.7198} & {\bf 0.7786}\\
 \hline
City &  PSNR & 25.04 & 25.59 & 25.80 & 26.15 & { 26.31} & 26.02 & {\bf 27.20}\\
 & SSIM & 0.5969 & 0.6520 & 0.6687 & 0.6947 & { 0.7218} & 0.6926 & {\bf 0.7818}\\
  \hline
Foliage &  PSNR& 23.62 & 24.44 & 24.62 & 24.95 & { 25.07} & 24.79 & {\bf 25.86}\\
 & SSIM & 0.5689 & 0.6451 & 0.6522 & 0.6731 & 0.7002 & 0.6639 & {\bf 0.7386}\\
   \hline
Walk & PSNR& 25.97 & 27.49 & 27.99 & 28.21 & 28.05 & { 28.42} & {\bf 29.55}\\
 & SSIM & 0.7957 & 0.8432 & 0.8584 & 0.8594 & 0.8583 & { 0.8705} & {\bf 0.8940}\\
\hline \hline
Average & PSNR & 23.78 & 24.70 & 25.02 & 25.31 & 25.39 & { 25.40} & {\bf 26.49}\\
 & SSIM & 0.6309 & 0.6948 & 0.7132 & 0.7282 & { 0.7464} & 0.7367 & {\bf 0.7982}\\
 \hline
\end{tabular}
\end{center}
\vspace{-2mm}
\caption{Comparison of PSNR values on the standard Vid4 dataset for scale factor 4. Best results are {\bf highlighted}.}
\label{tab:videox4}
\vspace{-1mm}
\end{table}

\begin{figure*}[]
\begin{center}
\begin{tabular}{cccccc}
          \animategraphics[trim = 220 170 450 270,width=0.15\textwidth,loop]{8}{"video/walk/bicubic/"}{1}{8} &
          \animategraphics[trim = 210 160 440 260,width=0.15\textwidth,loop]{8}{"video/walk/vespcn/"}{1}{8} &
          \animategraphics[trim = 220 170 450 270,width=0.15\textwidth,loop]{8}{"video/walk/temporal_dynamics/"}{1}{8} &
          \animategraphics[trim = 220 170 450 270,width=0.15\textwidth,loop]{8}{"video/walk/our_sisr/"}{1}{8} &                   
          \animategraphics[trim = 220 170 450 270,width=0.15\textwidth,loop]{8}{"video/walk/our_misr/"}{1}{8} &                              \animategraphics[trim = 220 170 450 270,width=0.15\textwidth,loop]{8}{"video/walk/orig/"}{1}{8}          \\
   (a) & (b) & (c) & (d) & (e) & (f)
\end{tabular}  
\vspace{-1mm}
   \caption{\label{fig:animatevideo} Comparison of output frames for video SR (x4) on a zoomed in region in the video $walk$ from vid4 dataset. The subfigures contain 8 consecutive frames from the results of (a) bicubic interpolation, (b) VESPCN, (c) Temp\_robust, (d) Our model (single-frame based), (d) Our model (multiple-frame based), and (f) ground truth. \textit{Videos can be
viewed by clicking on the images, when document is opened in Adobe Reader}.}
\end{center}
\vspace{-1.5em}
\end{figure*}

For higher scale factors e.g. 4, SISR becomes highly ill-posed and availability of neighboring LR frames to the network becomes crucial. Hence, we have provided qualitative comparisons between prior works, our SISR network and our multi-frame based network for $\times4$ video SR task in Fig. \ref{fig:video}. As can be observed from the $calendar$ example (first row), our multi-frame approach is able to produce the best results, wherein the readability of letters has further improved as compared to the existing methods and our single-frame based approach. For the frame of $walk$ video, one can note that both versions of our network are able to maintain the sharpness of the vertical line much better than the competing approaches. In addition, Fig. \ref{fig:animatevideo} shows 8 consecutive output frames for each method. (Please click on the figures to play the videos). The results of VESPCN and Temp\_robust partially suffer from distorted edges, missing texture and temporal fluctuations. Our SISR model leads to sharper results with fewer distortions but is limited in its capability to maintain temporal smoothness. In contrast, The frames estimated by our multi-frame based network are sharper, contain minimum fluctuations and are qualitatively consistent with the ground-truth video. These improvements are quantitatively reflected in Table \ref{tab:videox4}, where our method scores atleast 1 dB higher than all prior works, including our SISR model. Our network's superior performance demonstrates its capability of implicitly handing the temporal shifts and extracting HR information that is distributed across multiple LR frames.

\section{Conclusions}
\label{sec:summary}
We proposed a novel deep learning based approach for single image super-resolution. Our single-image SR network MDCN has been designed by effectively  combining the residual and dense connections. The effective
combination assists in better information flow through the network layers by reducing the redundant features and gradient vanishing problem. The scale recurrent framework has yielded better performance for higher scale factors with less number of parameters. The robustness of the network has been demonstrated for different scale factors on different datasets. The conventional pixel level
loss functions in the network fails to produce perceptually better results.
We included VGG loss as well as GAN based loss to generate photo-realistic
results. Different weights to those loses synthesized different results
that follow the perception-distortion trade-off. Further, the proposed MDCN structure has been utilized for video SR as a multi-frame restoration model. The resultant videos demonstrated the capability of our network in producing spatio-temporally consistent HR frames. \\
Refined and complete version of this work appeared in Neurocomputing 2019.

\bibliographystyle{elsarticle-num}
 \bibliography{deep_sr}

\end{document}